\documentclass[10pt,conference,a4paper]{IEEEtran}
\IEEEoverridecommandlockouts

\usepackage{mathtools}
\usepackage[utf8x]{inputenc}

\usepackage{subfigure}
\usepackage{moreverb}
\usepackage{epsfig}
\usepackage{amsmath,amssymb,bm,amsthm,mathrsfs,dsfont}
\usepackage{fancyhdr}
\usepackage{adjustbox,lipsum}
\usepackage[font={small}]{caption}
\usepackage{color,soul}
\usepackage{graphicx}
\usepackage{amsfonts}
\usepackage{cite}
\usepackage{microtype}

\usepackage{algorithmicx}
\usepackage{algpseudocode}
\usepackage[norelsize, linesnumbered, ruled, vlined, lined, boxed, commentsnumbered]{algorithm2e}

\enlargethispage{2\baselineskip}    

\allowdisplaybreaks
\usepackage[nodisplayskipstretch]{setspace} 
\usepackage{amsmath,amssymb,amsfonts}
\usepackage{graphicx}
\usepackage{booktabs}
\usepackage{multirow}
\usepackage{cite}
\usepackage{xcolor}

\usepackage{comment}
\usepackage{balance}

\begin{document}
\title{ 
Resource-Efficient Wi-Fi CSI-Based Sensing via Exploiting the Age of Samples
}
\author{\IEEEauthorblockN{
 Abolfazl Zakeri\IEEEauthorrefmark{1}, Nhan~Thanh~Nguyen\IEEEauthorrefmark{1},
 and Markku Juntti\IEEEauthorrefmark{1}}
 \vspace{-1 em }
 \\
 \IEEEauthorrefmark{1}\normalsize 
CWC-RT, University of Oulu, Finland,
 Email: 
\{abolfazl.zakeri,\,nhan.nguyen,\,markku.juntti\}@oulu.fi
\vspace{-1 em}
\\\thanks{This work was supported by the Research Council of Finland through 6G Flagship Program (no. 369116) and projects  DIRECTION (no. 354901) and DYNAMICS (no. 367702).}
}
\vspace{-2 em}
\maketitle
\vspace{-4 em}
\begin{abstract}
Wi-Fi channel state information (CSI)-based sensing must coexist with data communications, limiting the availability of temporally-dense CSI measurements. We formulate CSI-based human activity and identity recognition under an average sensing budget that limits the fraction of CSI measurement and reporting opportunities within a sensing session. The budget captures sensing–communication resource sharing, packet loss, and traffic-induced irregularity, which we model using deterministic (accumulated) and stochastic (Bernoulli) sampling policies. We propose a low-cost, age-aware WiFi sensing framework that encodes the age of each retained CSI sample and multiplicatively fuses it with the CSI embedding. 
On the NTU-Fi human activity recognition and person identification datasets, the proposed model outperforms both a CSI-only baseline and the time-aware attention model of the UniFi benchmark across most operating regimes. For person identification, it improves over UniFi by more than 10 percentage points, with the largest gains under strict sensing budgets.
\end{abstract}

\section{Introduction}\label{sec:intro}
Wi-Fi channel state information (CSI) sensing infers human activity and identity from the perturbations that body motion (and shape) induces in the propagating radio channel, without any worn or carried device~\cite{yousefi2017survey,ahmad2024survey}. Each received packet provides amplitude and phase responses across antennas and subcarriers, while a temporal sequence of such CSI frames captures the signature of the underlying motion. Deep learning models combining a convolutional encoder with a temporal aggregator have become the dominant approach for this task, and benchmarks such as SenseFi~\cite{yang2023sensefi} report top-1 accuracy above $90\%$ for activity recognition and person identification under \textit{dense, regular} CSI sampling.

However, such accuracy relies on regular, high-rate CSI sampling, which is rarely available when sensing coexists with communication. With the standardization of integrated sensing and communication (ISAC) in IEEE~802.11bf~\cite{du2025overview,meneghello2023isac}, sensing shares the medium with data traffic, and the sensing measurement rate is negotiated at the medium access control (MAC) layer to protect communication throughput~\cite{keshtiarast2025coexistence}; thus, the sensing rate reflects the sensing–communication resource share. Beyond this scheduled sparsity, packet loss and random traffic further reduce the number of available CSI frames and make their arrival times non-uniform. A sensing model that performs well only at high, regular sensing rates therefore has limited practical value

Prior work on sparse or irregular CSI follows three directions. Reconstruction methods, such as WiImg2.0~\cite{wiimg2024} and CSI-BERT~\cite{csibert}, recover dense CSI from sparse or lossy observations before inference, at the cost of additional computation, latency, and potential reconstruction errors. Resource-selection methods, such as Slim-Sense~\cite{singh2025slimsense}, prune subcarriers and antennas but do not address the temporal sensing rate. Closest to our setting, UniFi~\cite{dong2025unifi} performs sensing on irregularly sampled CSI drawn \textit{only} from ambient communication traffic, embedding absolute packet timestamps into a time-aware attention network that maps the samples onto a regular reference grid. However, none of these works evaluates inference accuracy explicitly as a function of an average sensing budget, nor do they treat the time elapsed since a sample was captured, i.e., its age, as an explicit descriptor of freshness. This notion is closely related to the age of information (AoI)~\cite{Roy_Survey}, which is widely used to optimize scheduling in communication networks but has not yet been explored as a trainable input feature for deep learning models in Wi-Fi sensing.

In this paper, we study Wi-Fi CSI-based sensing under an average sensing budget that constrains the fraction of sensing measurement instances yielding fresh CSI measurements. To emulate practical conditions, we consider two sampling algorithms: a deterministic (accumulated) policy modeling scheduled sparse sensing and a stochastic Bernoulli policy modeling random packet loss and traffic-driven irregularity. To handle the resulting sparse and non-uniform samples, we propose a low-cost, age-aware Wi-Fi sensing framework that encodes the age of each retained CSI sample and combines it with the CSI embedding via lightweight multiplicative fusion. The age provides explicit temporal context with negligible parameter overhead and is computed locally from the indices of the received measurements. The framework adapts the AoI-aware design we developed for sensing-aided beam prediction~\cite{zakeri_icc26} to Wi-Fi CSI sensing.

We evaluate the framework through extensive Monte Carlo experiments on the real-world NTU-Fi human activity recognition and person identification datasets, comparing it with a CSI-only baseline and the time-aware attention model of UniFi~\cite{dong2025unifi}. The proposed model outperforms both in most operating regimes, achieving an improvement of more than $10$ percentage points over UniFi for person identification.
\section{System Model and Problem Formulation}\label{sec:system}
This section describes the CSI measurement setup, the sensing budget and the CSI sampling algorithms, and the resulting inference task.
\\\indent 
We consider a single Wi-Fi sensing link with $N$ receive antennas and $K$
subcarriers, operating within an IEEE~802.11bf sensing measurement
session~\cite{chen2023wifi}. We model the session as a sequence of $T$
\emph{sensing measurement instances} (hereafter, \emph{instances}), indexed by
$t \in \{0, \dots, T-1\}$ and nominally spaced by $\Delta\tau$, where each
instance is an opportunity for one sounding exchange, e.g., a null data packet
(NDP) transmission.\footnote{This abstraction captures the essence of
sensing--communication resource sharing; protocol-level IEEE~802.11bf modeling
and evaluation in operational networks are left for future work. Like most
public WiFi sensing datasets, the data used in Section~\ref{sec:experiments}
were collected without coexisting data traffic; we therefore emulate the
resulting sparsity by subsampling dense CSI traces.}
Depending on the medium access state and whether a valid sounding packet is
received, a measurement may or may not take place at a given instance. When a
measurement occurs, the receiver processes the sounding signal to obtain a
single CSI snapshot, denoted as the CSI frame~$\mathbf{X}(t)$\footnote{Only the CSI amplitude is used in this work, as CSI phase measurements on commodity hardware are severely corrupted by carrier frequency offset, sampling frequency offset, and packet detection delay, which introduce large time-varying phase shifts that require strict hardware synchronization and signal processing to remove~\cite{ratnam2024preproc,yousefi2017survey}.}
\begin{equation}
    \mathbf{X}(t)=[X_{n,k}(t)] \in \mathbb{R}^{N\times K}, \qquad
    X_{n,k}(t) = \big| H_{n,k}(t) \big|,
    \label{eq:csi_frame}
\end{equation}
where the channel response of antenna $n$ at subcarrier $k$ is the
superposition of $L$ propagation paths,
\begin{equation}
    H_{n,k}(t) = \sum_{l=1}^{L}
        \eta_{n,l}(t)\, e^{-j2\pi\left(f_c + k\Delta f\right)\tau_l(t)} ,
    \label{eq:multipath}
\end{equation}
with $f_c$ the carrier frequency, $\Delta f$ the subcarrier spacing, and
$\eta_{n,l}(t)$, $\tau_l(t)$ the complex gain and delay of path $l$. Human motion
alters the body-reflected paths, so the temporal evolution of $\mathbf{X}(t)$ across the measurement session carries the sensing signal.

\subsection{Average Sensing Budget}
Under IEEE~802.11bf, sensing and data communication share the same medium, and the
rate of sensing measurements is bounded at the MAC layer so that sensing does not
degrade communication throughput beyond a negotiated level~\cite{keshtiarast2025coexistence}.
The fraction of instances devoted to sensing thus reflects the share of radio resources
assigned to sensing rather than to communication. Packet loss and traffic-driven irregularity further reduce the number of usable frames and make their arrival times
uneven. We capture these effects through an average sensing budget that limits how
often fresh CSI is available within each measurement session.
Therefore, better sensing performance under strict sensing rate limits indicates more resource-efficient Wi-Fi sensing algorithms.

Let $\beta(t)\!\in\!\{0,1\}$ be a binary indicator, where $\beta(t)\!=\!1$ if a
CSI measurement is acquired and reported at instance $t$ and $\beta(t)\!=\!0$ otherwise. We set
$\beta(0)\!=\!1$ to ensure that each measurement session contains at least one
CSI sample. We denote by $M\!:=\!\sum_{t=0}^{T-1}\beta(t)$ the number of sensed
frames and by $t_1\!<\!t_2\!<\!\cdots\!<\!t_M$ their instance indices. Since $M$
may be random, the average sensing budget constraint is
\begin{equation}
    \frac{1}{T}\,\mathbb{E}\left[M\right] \;\le\; \alpha,
    \qquad \alpha \in (0,1],
    \label{eq:budget}
\end{equation}
where the expectation is over the realizations of the sampling algorithm; for a
deterministic algorithm, \eqref{eq:budget} holds in every~session.

The budget $\alpha$ is set by the sensing--communication resource share. In
IEEE~802.11bf, the occurrence of sensing measurement instances is negotiated
during the sensing measurement setup~\cite{chen2023wifi}. For the nominal
instance spacing $\Delta\tau$, $\alpha$ bounds the mean sensing rate as
$f_{\mathsf{s}} \le \alpha/\Delta\tau$ and, for a fixed airtime per sounding
exchange, scales the airtime consumed by sounding proportionally. For instance,
a smaller $\alpha$ may be configured during peak data traffic hours, leaving a
larger fraction of the medium available for data communication.
Beyond dedicated sensing scheduling, $\alpha$ also captures packet losses or
irregular ambient traffic; it essentially models the effective fraction of
instances in which a usable CSI sample is collected.

\textit{Sampling Algorithms:}
We consider two sampling algorithms that realize the budget and represent two
regimes of practical interest: accumulated (Bresenham) sampling and Bernoulli
sampling.
The accumulated algorithm is a deterministic policy that spreads the $M$ sensed
instances near-uniformly over the measurement session: an accumulator increases by
$\alpha$ at each instance and triggers a measurement when it reaches unity,
producing evenly spaced samples that model scheduled dedicated sensing. Together
with $\beta(0)\!=\!1$, it yields $M\!=\!1+\lfloor (T\!-\!1)\alpha \rfloor$,
which equals $\alpha T$ whenever $\alpha T$ is an integer, so that
\eqref{eq:budget} holds with equality.
On the other hand, the Bernoulli algorithm is a stochastic policy in which $\beta(t)$ is drawn
independently as $\mathrm{Bernoulli}(\alpha)$ for $1\!\le\!t\!\le\!T\!-\!1$,
modeling random packet loss and traffic-driven irregular sensing. Hence,
$M\!=\!1+M'$ with $M'\!\sim\!\mathrm{Binomial}(T-1,\alpha)$, and the expected
sensing rate is $\mathbb{E}[M]/T = \alpha + (1-\alpha)/T$. That is, the budget
is met on average rather than strictly per session, and the forced sample at $t\!=\!0$
exceeds the nominal budget by less than $1/T$ 
For both algorithms, we refer to $\alpha$ as the nominal
sensing budget.

\subsection{Measurement Session and Data Staleness}
The $T$ consecutive instances form one \emph{measurement session}, over which the a machine learning model
produces a single decision evaluated at the final instance $t\!=\!T\!-\!1$. Within a
session, the sampling decisions $\{\beta(t)\}_{t=0}^{T-1}$ determine which frames are
available: the sensed instances are denoted by $\{t_1,\dots,t_M\}=\{t:\beta(t)=1\}$. These sampling times, in turn, determine the staleness of the information at the decision instant. We quantify this
staleness by AoI~\cite{Roy_Survey}. The age of the $m$-th
sensed frame, defined relative to the session decision instant, i.e., $t=T-1$, is
\begin{equation}
    \delta(t_m) = (T-1) - t_m, \qquad m = 1,\dots,M ,
    \label{eq:age}
\end{equation}
which is determined entirely by the sampling decisions $\beta(\cdot)$ and
corresponds to the physical age $\delta(t_m)\Delta\tau$. The most recent sensed frame relative to the decision time has the smallest age, and earlier sensed frames have larger ages,
denoting older and less fresh measurements. Unsensed instances produce no frame and hence no age, so only the ages of the $M$ sensed frames form the observation samples.
Since the receiver knows at which instances measurements were obtained, these
ages can be computed locally from the instance indices (or measurement
timestamps), with negligible computation and no additional~signaling.

\subsection{Packed Observation and Machine Learning Task}
Rather than carrying stale frames forward into unsensed instances, we adopt a
\emph{packed} representation that retains only the $M$ sensed frames, each tagged with
its age:
\begin{equation}
    \mathcal{X} = \big\{\, \big(\mathbf{X}(t_m),\, \delta(t_m) \big) \big\}_{m=1}^{M}.
    \label{eq:packed}
\end{equation}
Packing removes the unsensed instances, reduces the encoder computation in
proportion to $M$, i.e., to $\alpha$ on average, and makes the age the explicit
carrier of each frame's temporal~position.

Let $\mathcal{Y}$ denote the label set corresponding to the human activity classes or the identities to be recognized.
The machine learning (inference) task is to learn a classifier
$f_{\boldsymbol{\theta}}:\mathcal{X}\!\to\!\mathcal{Y}$ that minimizes the expected classification
error under a given budget and sampling algorithm.
With a dataset of $I$ sessions $\{(\mathcal{X}_i,\, y_i)\}_{i=1}^{I}$, the
training objective is the empirical loss
\begin{align}
\underset{\boldsymbol{\theta}}{\mbox{minimize}}\;~~
    \dfrac{1}{I}\sum_{i=1}^{I}
    \mathcal{L}\big(f_{\boldsymbol{\theta}}(\mathcal{X}_i),\, y^*_i\big),
    \label{eq:objective}
\end{align}
where $\mathcal{X}_i$ is the packed observation of session $i$, $y^*_i \in
\mathcal{Y}$ is its true label, and $\mathcal{L}$ denotes the cross-entropy function given by
\begin{align}
           \mathcal{L}\big(f_{\boldsymbol{\theta}}(\mathcal{X}),\, {y^*}\big)
= -\log\!\left(
   \frac{\exp\!\big(f_{\boldsymbol{\theta},y^*}(\mathcal{X})\big)}
        {\sum_{y=1}^{|\mathcal{Y}|}\exp\!\big(f_{\boldsymbol{\theta},y}(\mathcal{X})\big)}
  \right).
\end{align}
The age values $\delta(t_m)$ are encoded and fed to the model as described in the next section.

\section{Age-Aware Wi-Fi Sensing Framework} \label{sec:framework}
This section presents the proposed age-aware sensing framework. We first give an overview of how the budget-constrained packed observations are processed to
produce a classification decision, and then detail each component: the CSI
encoder, the age encoder, their multiplicative fusion, and the classifier,
followed by the training procedure.

\subsection{Overview}
\begin{figure*}[t]
  \centering
  \includegraphics[width=0.86\textwidth]{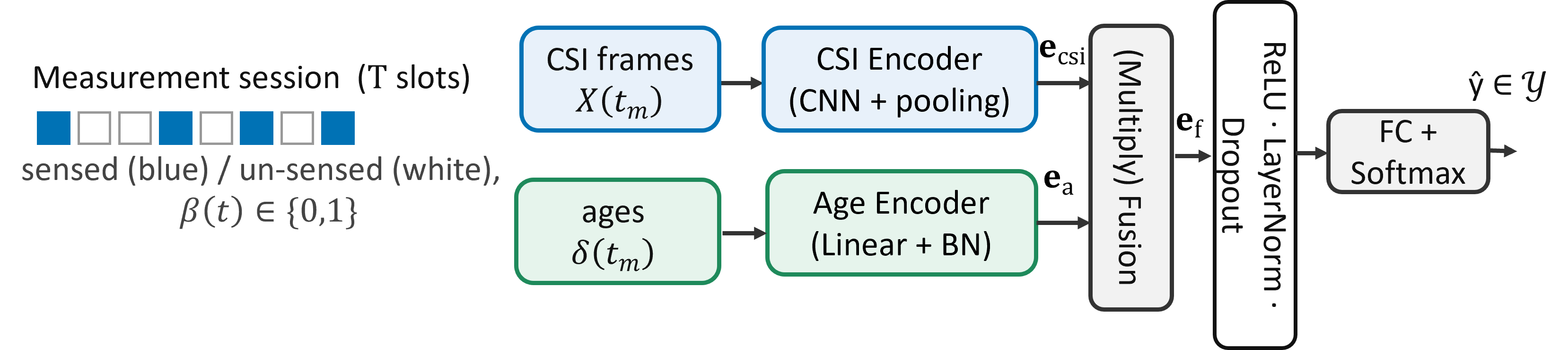}
  \caption{Proposed age-aware Wi-Fi CSI sensing framework. The packed sensed frames and
  their ages are encoded by two different encoders; the age embedding $\mathbf{e}_{\mathsf{a}}\!\in\!(0,1)^{d}$ and
   the CSI embedding $\mathbf{e}_{\mathsf{csi}}$ are fused by element-wise multiplication to form the fused embedding $\mathbf{e}_{\mathsf{f}}$ before~classification.}
  \label{fig_framework}
  \vspace{-2 em}
\end{figure*}

The framework, shown in Fig.~\ref{fig_framework}, takes the packed observation
$\mathcal{X}=\{(\mathbf{X}(t_m),\delta(t_m))\}_{m=1}^{M}$ of a measurement session and
produces one class decision. It has two parallel branches. The CSI branch encodes the
$M$ sensed frames into a single embedding $\mathbf{e}_{\mathsf{csi}}\!\in\!\mathbb{R}^{d}$, with $d$ being the embedding dimension, that summarizes the motion content of the session. The age branch encodes the
corresponding ages into an embedding
$\mathbf{e}_{\mathsf{a}}\!\in\!(0,1)^{d}$ that acts as a per-feature weight
conditioned on the staleness of the session. The two embeddings are fused
by an element-wise product, refined by a small post-fusion block, and mapped to class
scores. The age branch carries no environmental information of its own; it acts as a
freshness-dependent gate on the CSI features, and it requires no additional sensing
or signaling, since the ages follow from the instance indices of the received
frames (Section~\ref{sec:system}). The design adds negligible parameters over a
plain CSI classifier, which is essential for an on-device sensing model.
\subsection{CSI Encoder}
The CSI encoder maps the packed CSI tensor
$[\,N,K,M\,]$ to the embedding $\mathbf{e}_{\mathsf{csi}}$. The
$N$ antenna streams form the input channels, the $K$ subcarriers the
height, and the $M$ packed frames the width, so that the convolutions operate jointly
over the subcarrier and temporal axes. Three convolutional blocks
($N\!\to\!32$ with a $15\!\times\!7$ kernel and stride $9\!\times\!1$,
$32\!\to\!64$ with $3\!\times\!3$, and $64\!\to\!96$ with $7\!\times\!3$), each followed
by a ReLU, extract spatial-spectral features. An adaptive average pooling layer
collapses the result to a fixed $96\!\times\!4$ map, which is flattened and projected
by a linear layer with ReLU to $\mathbf{e}_{\mathsf{csi}}\!\in\!\mathbb{R}^{d}$ with
$d\!=\!128$. Because the pooling is adaptive, the encoder accepts any number of packed
frames $M$ and yields a fixed-size embedding, so a single architecture serves all
budgets. The temporal aggregation of the session is therefore performed inside the
encoder by pooling over the packed-frame axis, without any recurrent module.
\subsection{Age Encoder}
Each age $\delta(t_m)$ is a scalar and cannot be combined with the high-dimensional CSI
embedding directly. The age encoder takes the normalized age
$\bar{\delta}(t_m)\!:=\!\delta(t_m)/(T\!-\!1)\!\in\![0,1]$, lifts it to the
embedding dimension through a shared linear layer, batch normalization, and a
sigmoid, and then averages over the $M$ frames:
\begin{equation}
    \mathbf{e}_{\mathsf{a}}
    = \frac{1}{M}\sum_{m=1}^{M}
      \sigma\!\big(\mathsf{BN}(\mathbf{w}_{\mathsf{a}}\,\bar{\delta}(t_m)+\mathbf{b}_{\mathsf{a}})\big)
    \in (0,1)^{d},
    \label{eq:age_enc}
\end{equation}
where $\mathbf{w}_{\mathsf{a}}, \mathbf{b}_{\mathsf{a}}\in \mathbb{R}^{d}$ are the
weights and biases of the linear layer, $\sigma(\cdot)$ is the sigmoid function,
and $\mathsf{BN}$ denotes batch normalization with trainable scale and shift
parameters. Since $\mathsf{BN}$ directly follows the linear layer, the encoder
output is invariant to any positive rescaling of the age; hence, using the
physical age $\delta(t_m)\Delta\tau$ instead of $\bar{\delta}(t_m)$ is equivalent.
The sigmoid constrains each dimension of $\mathbf{e}_{\mathsf{a}}$ to
$(0,1)$, so the embedding can be read as a per-dimension weight, and averaging over
the frames yields a single session-level freshness descriptor aligned
with $\mathbf{e}_{\mathsf{csi}}$. As the average in~\eqref{eq:age_enc} is
permutation-invariant, $\mathbf{e}_{\mathsf{a}}$ summarizes the sampling pattern
of the session, i.e., the number and spread of the sensed frames, while the order
of the frames is retained by the CSI encoder along the packed-frame axis.

\textit{Age Fusion:} We fuse the two branches by an element-wise (Hadamard) product,
\begin{equation}
    \mathbf{e}_{\mathsf{f}}
    = \mathbf{e}_{\mathsf{csi}} \odot \mathbf{e}_{\mathsf{a}}.
    \label{eq:fusion}
\end{equation}
We opt for multiplicative fusion because the ages describe the reliability of the
CSI rather than an independent modality: the product in~\eqref{eq:fusion} lets
$\mathbf{e}_{\mathsf{a}}\!\in\!(0,1)^{d}$ scale each CSI feature dimension by a
learned, freshness-dependent factor, whereas additive or concatenation fusion would
inject staleness as a separate feature. In a related study on age-aware
sensing-aided beam prediction~\cite{zakeri2026freshness}, multiplicative fusion
outperformed additive and concatenation-based fusion; a comparison of fusion
operators for Wi-Fi CSI sensing is left for future work.\footnote{Since the ages in our emulation lie on a regular instance grid, they are common to all sessions at $\alpha\!=\!1$ and under accumulated sampling, where the age branch
acts as a learned, budget-specific feature scaling; in practice, contention and
packet losses make measurement times, and thus ages, irregular across sessions, a
regime emulated by the Bernoulli algorithm.
}

\textit{Post-Fusion Block and Classifier:} The fused embedding is refined by a ReLU, layer normalization, and dropout, and then
mapped to class scores by a linear classifier:
\begin{align}
    \hat{\mathbf{e}} &= \mathrm{Dropout}\big(\mathrm{LayerNorm}(\mathrm{ReLU}(\mathbf{e}_{\mathsf{f}}))\big), \\
    \hat{\mathbf{y}} &= \mathbf{W}_{\mathsf{c}}\,\hat{\mathbf{e}} + \mathbf{b}_{\mathsf{c}} \in \mathbb{R}^{|\mathcal{Y}|}.
\end{align}
The age encoder and its batch normalization introduce $4d$ parameters and the layer
normalization another $2d$, i.e., ${6d\!=\!768}$ parameters at $d\!=\!128$ in addition
to the CSI-only model (e.g., $208.6$k vs.\ $207.8$k in total for NTU-Fi HAR). The added
computation is $\mathcal{O}(Md)$, which is negligible relative to the CSI encoder,
whose cost grows linearly with $M$.
The proposed model therefore achieves age-aware inference at essentially the same complexity as the CSI-only model.

\textit{Variable-Length Sessions:} Under Bernoulli sampling the number of sensed frames $M$ varies across sessions. Each
training batch is padded to its largest $M$, and the adaptive pooling in the CSI encoder
and the averaging in~\eqref{eq:age_enc} are replaced by length-aware masked operations,
so that padded positions are excluded from the temporal pooling and the age averaging.
Under accumulated sampling $M$ is fixed by $\alpha$ and no masking is~required.

\textit{Training Procedure:} The model is trained end-to-end by minimizing the cross-entropy
loss~\eqref{eq:objective} with the Adam optimizer. Under accumulated sampling, the mask
is deterministic and shared across epochs; under Bernoulli sampling a fresh mask is drawn
for each training sample at every epoch, which exposes the model to many frame placements
and acts as temporal augmentation, while validation and test masks are fixed for
reproducibility. The complete procedure is given in~Algorithm~\ref{alg:train}.

\begin{algorithm}[t]
\SetAlFnt{\small}            
\SetAlCapFnt{\small}         
\SetInd{0.3em}{0.6em}        
\SetAlgoSkip{}               
\setlength{\algomargin}{0.5em}
\caption{Age-Aware Wi-Fi CSI-based Sensing}
\label{alg:train}
\KwIn{training set $\mathcal{D}$, nominal budget $\alpha$, sampling algorithm $\mathcal{S}\!\in\!\{\text{Accumulated},\text{Bernoulli}\}$, number of instances~$T$}
\KwOut{trained parameters $\boldsymbol{\theta}$}
initialize $\boldsymbol{\theta}$\;
\For{each epoch}{
  \For{each session $(\{\mathbf{X}(t)\}_{t=0}^{T-1},y)\in\mathcal{D}$}{
    draw mask $\beta(\cdot)$ from $\mathcal{S}$ with budget $\alpha$\tcp*{fresh per epoch if sampling is Bernoulli}
    $\{t_m\}\!\leftarrow\!\{t:\beta(t)\!=\!1\}$;\quad $\delta(t_m)\!\leftarrow\!(T\!-\!1)-t_m$\;
    pack $\mathcal{X}\!\leftarrow\!\{(\mathbf{X}(t_m),\delta(t_m))\}_{m=1}^{M}$\;
    $\mathbf{e}_{\mathsf{csi}}\!\leftarrow\!\mathrm{CSIEnc}(\{\mathbf{X}(t_m)\})$\;
    $\mathbf{e}_{\mathsf{a}}\!\leftarrow\!\mathrm{AgeEnc}(\{\delta(t_m)\})$ \tcp*{Eq.~\eqref{eq:age_enc}}
    $\mathbf{e}_{\mathsf{f}}\!\leftarrow\!\mathbf{e}_{\mathsf{csi}}\odot\mathbf{e}_{\mathsf{a}}$\;
    $\hat{\mathbf{e}}\!\leftarrow\!\mathrm{Dropout}(\mathrm{LayerNorm}(\mathrm{ReLU}(\mathbf{e}_{\mathsf{f}})))$\;
    $\hat{\mathbf{y}}\!\leftarrow\!\mathbf{W}_{\mathsf{c}}\,\hat{\mathbf{e}} + \mathbf{b}_{\mathsf{c}}$\;
    update $\boldsymbol{\theta}$ by Adam on $\mathcal{L}(\hat{\mathbf{y}},y^*)$\;
  }
  early-stop on validation loss\;
}
\Return $\boldsymbol{\theta}$\;
\end{algorithm}
\vspace{-1 em}
\section{Simulation Results}\label{sec:experiments}
This section evaluates the proposed age-aware WiFi sensing model against four benchmarks on
two NTU-Fi tasks, under two temporal sampling algorithms, across the nominal sensing budget
range $\alpha \in (0,1]$. We first describe our training and test datasets construction, preprocessing, and benchmarks, then report top-1 accuracy as
a function of $\alpha$.
\subsection{Dataset Construction}
We use two tasks from the Nanyang Technological University WiFi (NTU-Fi) collection of the SenseFi benchmark~\cite{},
both acquired with an Atheros AR9580 NIC at $\approx\!1000$~Hz over $K\!=\!114$
subcarriers and $N=3$ antenna streams. NTU-Fi Human Activity Recognition \textbf{(NTU-Fi HAR)} is a $6$-class
activity-recognition task (box, circle, clean, fall, run, walk) with $936$
training and $264$ test samples. \textbf{NTU-Fi HumanID} is a $14$-class
gait-based person-identification task with $546$ training and $294$ test samples.
HumanID is the more temporally sensitive task, as gait discrimination requires resolving periodic patterns that span the observation window.

\textit{Preprocessing:} An identical SenseFi pipeline is applied to every sample before temporal sampling: the raw amplitude tensor
$\mathrm{CSI}_{\mathsf{amp}}\!\in\!\mathbb{R}^{342\times2000}$ is standardized with
fixed dataset-level scalar statistics $(\mu,\sigma)$ adopted from SenseFi~\cite{yang2023sensefi},
downsampled $4\times$ along time ($2000\!\to\!500$), and reshaped to $[3,114,500]$.
The statistics are $(\mu,\sigma)\!=\!(42.32,4.98)$ for HAR and $(38.83,5.97)$ for~HumanID;
as a single affine transform applied identically to all samples, this
standardization carries no sample- or label-specific information.
Accordingly, each measurement session comprises $T\!=\!500$ instances with nominal
spacing $\Delta\tau\!=\!4$~ms, i.e., a $2$~s observation window, so that the budget range
$\alpha\!\in\![0.01,1]$ corresponds to mean sensing rates of $2.5$--$250$~Hz.

NTU-Fi provides predefined training and test partitions. From the training
partition, we hold out $20\%$ as a validation set for early stopping, training on the remaining $80\%$; the test partition is the same as the original partition in NTU-Fi.
Each Monte-Carlo run re-draws the train/validation split and the data-loader
shuffle (and, under Bernoulli sampling, the validation and test masks) from an
independent seed. The runs are paired across methods: within a run, all compared
models use the same train/validation split and the same test sampling~masks.

\textit{Sampling Algorithms Realization:} The two sampling algorithms are defined in
Section~\ref{sec:system}; here we describe their experimental realization. 
Let us define the binary sequence $\{\beta(t)\}_{t=0}^{T-1}$ as the sampling
mask. Under accumulated sampling, this mask is deterministic given $\alpha$ and
is therefore identical across the training, validation, and test datasets. Under Bernoulli sampling, a fresh realization of $\{\beta(t)\}$ is drawn for
each training sample at every epoch, providing temporal augmentation, while the
validation and test realizations are fixed by per-run seeds for reproducibility. The variable
number of sensed frames under Bernoulli sampling is handled at batching time by
zero-padding each batch to its local maximum $M_{\max}$ and applying length-aware masked
mean pooling, which excludes the padded columns from the pooling operations.

Training hyperparameter settings are summarised in Table~\ref{tab:training}. All results are reported
as mean $\pm$ standard deviation over $50$ Monte-Carlo~runs.
\begin{table}[t]
\centering
\small
\caption{Training configuration per dataset.}
\label{tab:training}
\renewcommand{\arraystretch}{0.9}   
\begin{tabular}{lcc}
\toprule
Parameter & HAR & HumanID \\
\midrule
Batch size            & 64 & 32 \\
Optimiser             & \multicolumn{2}{c}{Adam, $\eta\!=\!10^{-3}$} \\
Weight decay          & $10^{-4}$ & $3\!\times\!10^{-4}$ \\
LR schedule    & \multicolumn{2}{c}{warmup 5 ep., then} \\
               & \multicolumn{2}{c}{cosine to $\eta_{\min}\!=\!10^{-6}$} \\
Max epochs            & 100/150$^\dagger$ & 150 \\
Early-stop patience   & 15 & 20 \\
Dropout               & 0.3 & 0.4 \\
MC runs               & 50 & 50 \\
Validation fraction   & \multicolumn{2}{c}{20\%} \\
\bottomrule
\end{tabular}
\\[2pt]
{\footnotesize $^\dagger$ Accumulated / Bernoulli.}
\end{table}
\subsection{Results and Discussion}
We consider the following scenarios and benchmarks and evaluate their performance.
\begin{itemize}
\item \textbf{CSI+Age} (proposed): per-step age encoding multiplicatively fused with the
CSI embedding, as shown in Fig.~\ref{fig_framework}, and detailed in Section~\ref{sec:framework}.
\item \textbf{CSI-only}: the same CSI encoder and linear classifier, without the age
branch and the post-fusion block;
\item \textbf{mTAN (UniFi)}: our reimplementation of the multi-time attention network
(mTAN) of the UniFi framework~\cite{dong2025unifi}, following its description, with
$Q\!=\!64$ reference query points, learnable sinusoidal time embeddings, single-head
cross-attention, and a post-attention gated recurrent unit (GRU). For a fair comparison,
it uses the same CNN as per-frame feature extractor and the training settings of~Table~\ref{tab:training}.
\end{itemize}
In addition to these benchmarks, we consider a baseline, termed \emph{Last-$M$},
that retains only the $M$ most recent frames of the session instead of the $M$ frames
selected by the sampling algorithms. It uses the same budget $M$ but concentrates it
into a short recent burst rather than spreading it across the session. Hence, comparing it
against the accumulated configurations at equal $M$ isolates the effect of temporal
spread from the number of frames. Since a most-recent-$M$ truncation is not meaningful
once the sensed instances are randomly placed, Last-$M$ is evaluated under the accumulated algorithm
only. To also examine the impact of age encoding within this baseline, we include two
variants:
\begin{itemize}
\item \textbf{Last-$M$ CSI-only}: CSI-only on the $M$ most recent~frames, i.e., ${\{t=T-M,\dots, T-1\}}$; 
\item \textbf{Last-$M$ CSI+Age}: the proposed fusion on the $M$ most recent frames; this examines whether the age branch compensates for poor temporal coverage.
\end{itemize}

\begin{figure}[t]
\centering
\subfigure[Accumulated sampling]{%
  \includegraphics[width=0.48\columnwidth]{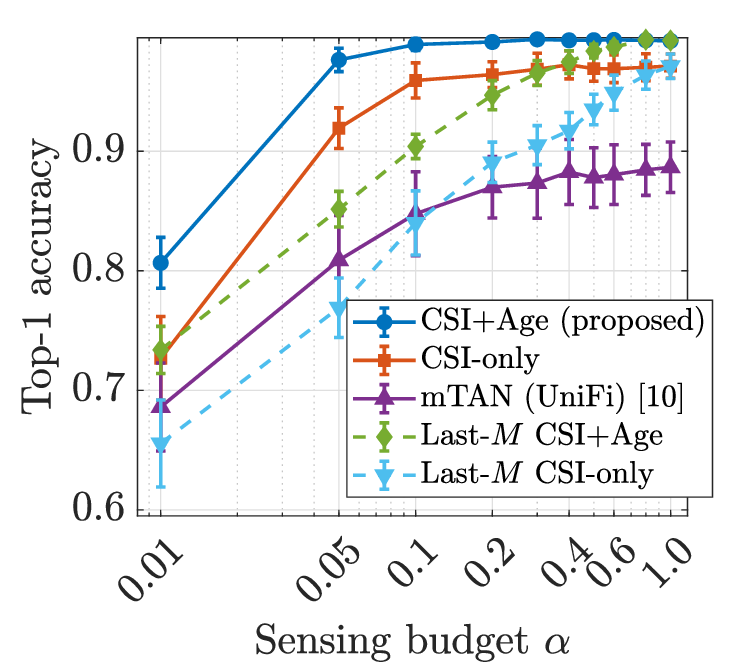}%
  \label{fig_humanid_brs}%
}%
\hspace{-2mm}%
\subfigure[Bernoulli sampling]{%
  \includegraphics[width=0.48\columnwidth]{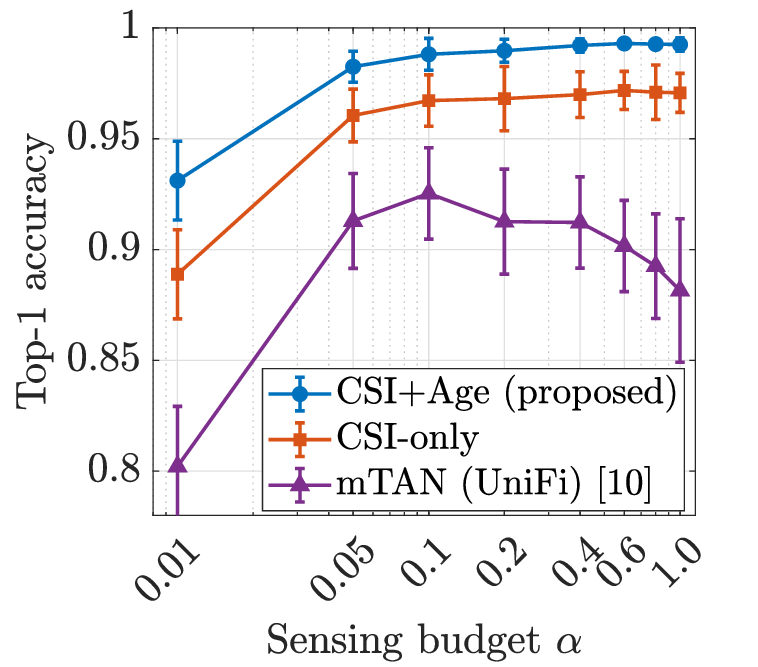}%
  \label{fig_humanID_bern}%
}
\caption{Top-1 accuracy vs.\ nominal sensing budget $\alpha$ (log scale) on NTU-Fi HumanID.}
\label{fig_humanID_}
\vspace{-1em}
\end{figure}

Fig.~\ref{fig_humanID_} shows the top-1 accuracy on the NTU-Fi HumanID dataset as a
function of the sensing budget $\alpha$ for the compared algorithms; note that the Last-$M$ baseline is sampling-agnostic and is therefore plotted only in the accumulated
case of Fig.~\ref{fig_humanid_brs}. The proposed CSI+Age model is the most accurate
across nearly the entire budget range and degrades gracefully as $\alpha$ shrinks,
retaining $80.7\%$ at $\alpha\!=\!0.01$ under accumulated sampling where the
CSI-only model drops to $72.7\%$ and the UniFi benchmark to $68.6\%$.
The gain of CSI+Age over CSI-only is largest at tight budgets, $8.0$ percentage points
(pp) at $\alpha\!=\!0.01$, and narrows toward full sensing ($2.1$~pp at $\alpha\!=\!1$).
\\\indent 
From the figure, two structural effects are observed. First, at equal $M$, the accumulated CSI-only model outperforms Last-$M$ CSI-only by
$7.1$~pp at $\alpha\!=\!0.01$, confirming that covering the session matters more than merely using recent frames. Second, the UniFi benchmark trails the proposed model by a
wide margin throughout, exceeding $10$~pp at full sensing under both sampling algorithms.
Under Bernoulli sampling, it peaks at $\alpha\!=\!0.1$ and then declines, whereas CSI+Age
saturates at about $99.2\%$. A possible reason is that the fixed grid of $Q\!=\!64$
reference points compresses the $M$ frames into $Q$ attention outputs, which becomes a
bottleneck when $M\!\gg\!Q$, together with the weaker augmentation effect of random
masking as $\alpha\!\to\!1$. Person identification is noticeably more budget-sensitive than activity
recognition, which amplifies both the degradation at low budgets and the benefit of the proposed~model.

\begin{figure}[t]
\centering
\subfigure[Accumulated sampling]{%
  \includegraphics[width=0.48\columnwidth]{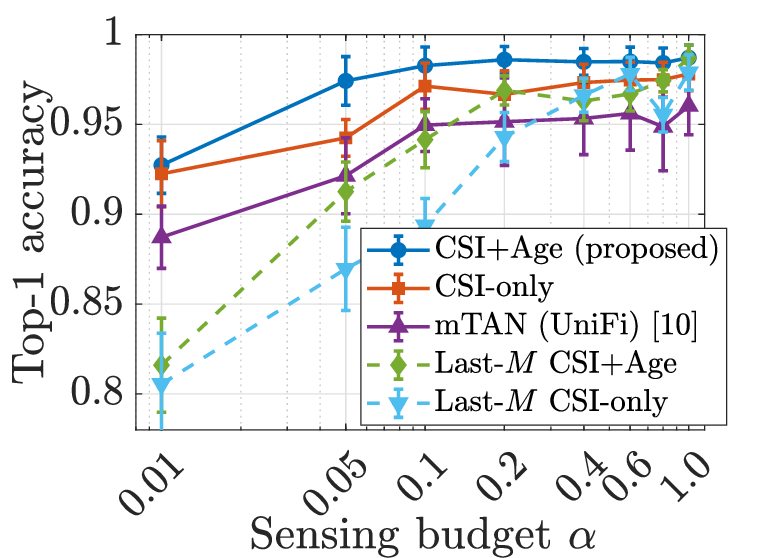}%
  \label{fig_har_brs}%
}%
\hspace{-2mm}%
\subfigure[Bernoulli sampling]{%
  \includegraphics[width=0.48\columnwidth]{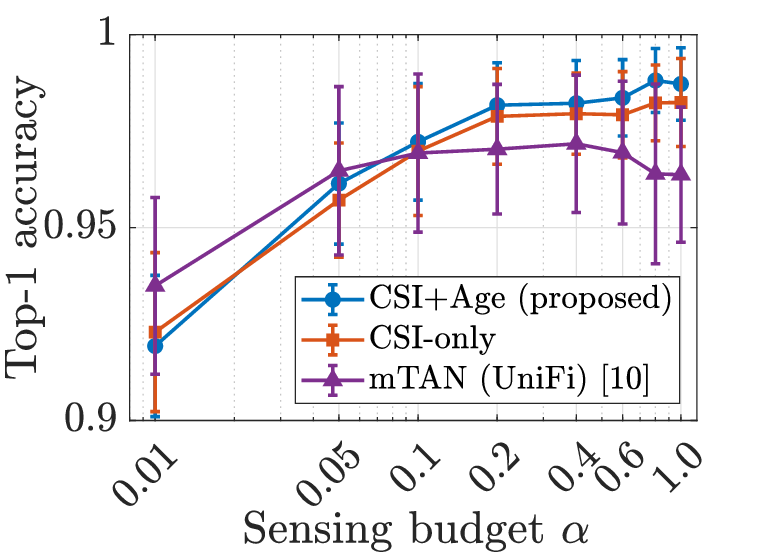}%
  \label{fig_har_bern}%
}
\caption{Top-1 accuracy vs.\ nominal sensing budget $\alpha$ (log scale) on NTU-Fi HAR. 
}
\label{fig_har}
\vspace{-1em}
\end{figure}

Fig.~\ref{fig_har} shows the top-1 accuracy on the NTU-Fi HAR dataset versus the sensing
budget $\alpha$. Under accumulated sampling in Fig.~\ref{fig_har_brs} the proposed CSI+Age
model is again the most accurate, or on par, at every budget, reaching $92.7\%$ at $\alpha\!=\!0.01$
and $98.7\%$ at full sensing, and the temporal-spread effect is pronounced: at equal
$M\!=\!5$ the accumulated CSI-only model attains $92.3\%$ against only $80.6\%$ for
Last-$M$ CSI-only, an $11.7$~pp gap that isolates session coverage from frame count. The
UniFi benchmark stays well below both CSI-based models across the range. Under Bernoulli sampling in Fig.~\ref{fig_har_bern} the three models are much closer, since the per-epoch
random masking already exposes the encoder to diverse frame placements and thus acts as
temporal augmentation; here, the UniFi benchmark is marginally best at the two tightest
budgets ($93.5\%$ at $\alpha\!=\!0.01$ and $96.5\%$ at $\alpha\!=\!0.05$) but then flattens
and declines with budget, while CSI+Age overtakes it from $\alpha\!\ge\!0.1$ onward and
rises to $98.8\%$. The age branch yields a small improvement over CSI-only for
$\alpha\!\ge\!0.05$ ($0.3$--$0.6$~pp, within one standard deviation). Overall, the gain on
HAR is smaller than on HumanID, consistent with activity recognition being less
temporally sensitive.

\section{Conclusion}\label{sec:conclusion}
We investigated the impact of limited and irregular CSI availability on Wi-Fi sensing and showed that exploiting sample age can improve recognition performance without substantially increasing model complexity. Across the evaluated sampling regimes, age awareness was most beneficial at low sensing rates and for person identification, where temporal information is more critical. The results also show that temporal coverage can be more important than sample recency: distributing a fixed number of samples across the measurement session generally provides greater benefit than concentrating them in a recent window. These findings highlight sample age as a simple and effective side information for robust Wi-Fi sensing under constrained CSI availability.
Future work will investigate alternative fusion mechanisms, additional datasets, and evaluation with CSI collected in operational IEEE~802.11bf networks.
\bibliographystyle{ieeetr}
\bibliography{Bib_References/conf_short,
Bib_References/IEEEabrv,
Bib_References/Bibliography, Bib_References/multimodalsensing_Bio, Bib_References/ML_Bio,Bib_References/wifisensing}

@ARTICLE{Roy_Survey,  author={R. D. {Yates} and Y. {Sun} and D. {Richard Brown} and S. K. {Kaul} and E. {Modiano} and S. {Ulukus}},  journal={IEEE J. Sel. Areas Commun.},   title={Age of Information: An Introduction and Survey},   year={May, 2021. },
volume={39},  number={5},  pages={1183-1210}, }

@string{ icc = {Proc. IEEE Int. Conf. Commun.}}

@string{ infocom = {Proc. IEEE Int. Conf. on Comp. Commun.}}

@string{ infocom = {Proc. IEEE Int. Conf. on Computer. Commun.  (INFOCOM)}}

@article{zakeri_icc26,
  title={{AoI}-Aware Machine Learning for Constrained Multimodal Sensing and Communications},
  author={A. Zakeri and Nguyen, Nhan Thanh and Alkhateeb, Ahmed and Juntti, Markku},
  journal=icc,
  year={Accepted, May 2026}
}

@article{du2025overview,
  author  = {Du, Rui and Hua, Haocheng and Xie, Hailiang and Song, Xianxin and Lyu, Zhonghao and Hu, Mengshi and Narengerile and Xin, Yan and McCann, Stephen and Montemurro, Michael and Han, Tony Xiao and Xu, Jie},
  title   = {An overview on {IEEE} 802.11bf: {WLAN} sensing},
  journal = {IEEE Commun. Surveys Tuts.},
  volume  = {27},
  number  = {1},
  pages   = {184--217},
  year    = {2025},
  doi     = {10.1109/COMST.2024.3408899}
}

@article{chen2023wifi,
  author  = {Chen, Cheng and Song, Hao and Li, Qinghua and Meneghello, Francesca and Restuccia, Francesco and Cordeiro, Carlos},
  title   = {{Wi-Fi} sensing based on {IEEE} 802.11bf},
  journal = {IEEE Commun. Mag.},
  volume  = {61},
  number  = {1},
  pages   = {121--127},
  month   = jan,
  year    = {2023},
  doi     = {10.1109/MCOM.007.2200347}
}

@article{meneghello2023isac,
  author  = {Meneghello, Francesca and Chen, Cheng and Cordeiro, Carlos and Restuccia, Francesco},
  title   = {Toward integrated sensing and communications in {IEEE} 802.11bf {Wi-Fi} networks},
  journal = {IEEE Commun. Mag.},
  volume  = {61},
  number  = {7},
  pages   = {128--133},
  month   = jul,
  year    = {2023},
  doi     = {10.1109/MCOM.003.2200563}
}

@article{yousefi2017survey,
  author  = {Yousefi, Siamak and Narui, Hirokazu and Dayal, Sankalp and Ermon, Stefano and Valaee, Shahrokh},
  title   = {A survey on behavior recognition using {WiFi} channel state information},
  journal = {IEEE Commun. Mag.},
  volume  = {55},
  number  = {10},
  pages   = {98--104},
  month   = oct,
  year    = {2017},
  doi     = {10.1109/MCOM.2017.1700082}
}

@article{ahmad2024survey,
  author  = {Ahmad, Iftikhar and Ullah, Arif and Choi, Wooyeol},
  title   = {{WiFi}-based human sensing with deep learning: Recent advances, challenges, and opportunities},
  journal = {IEEE Open J. Commun. Soc.},
  volume  = {5},
  pages   = {3595--3623},
  year    = {2024},
  doi     = {10.1109/OJCOMS.2024.3411529}
}

@article{keshtiarast2025coexistence,
  author  = {Keshtiarast, Navid and Bishoyi, Pradyumna Kumar and Lumbantobing, Ido Manuel and Petrova, Marina},
  title   = {When next-gen sensing meets legacy {Wi-Fi}: Performance analyses of {IEEE} 802.11bf and {IEEE} 802.11ax coexistence},
  journal = {IEEE Trans. Mobile Comput.},
  year    = {2026},
  doi     = {10.1109/TMC.2026.3730742},
  note    = {early access; arXiv:2503.04637}
}

@misc{dong2025unifi,
  author       = {Dong, Gaofeng and Yang, Kang and Srivastava, Mani},
  title        = {{UniFi}: Combining irregularly sampled {CSI} from diverse communication packets and frequency bands for {Wi-Fi} sensing},
  howpublished = {arXiv:2512.22143},
  year         = {2025}
}

@article{wiimg2024,
  author  = {Zheng, Xiaolong and Yang, Kangning and Xiong, Jie and Liu, Liang and Ma, Huadong},
  title   = {Pushing the limits of {WiFi} sensing with low transmission rates},
  journal = {IEEE Trans. Mobile Comput.},
  volume  = {23},
  number  = {11},
  pages   = {10265--10279},
  month   = nov,
  year    = {2024},
  doi     = {10.1109/TMC.2024.3374046}
}

@inproceedings{csibert,
  author    = {Zhao, Zijian and Chen, Tingwei and Meng, Fanyi and Li, Hang and Li, Xiaoyang and Zhu, Guangxu},
  title     = {Finding the missing data: A {BERT}-inspired approach against package loss in wireless sensing},
  booktitle = {Proc. IEEE Conf. Comput. Commun. Wkshps. (INFOCOM WKSHPS)},
  pages     = {1--6},
  year      = {2024},
  doi       = {10.1109/INFOCOMWKSHPS61880.2024.10620769}
}

@article{singh2025slimsense,
  author  = {Singh, Vishal Kumar and Walecha, Aditya and Gera, Aman and Jay, Rishabh and Bhattacharya, Arani and Maity, Mukulika},
  title   = {{Slim-Sense}: A resource efficient {Wi-Fi} sensing framework towards integrated sensing and communication},
  journal = {Proc. ACM Interact. Mob. Wearable Ubiquitous Technol.},
  volume  = {9},
  number  = {1},
  pages   = {17:1--17:33},
  year    = {2025},
  doi     = {10.1145/3712271}
}

@article{ratnam2024preproc,
  author  = {Ratnam, Vishnu V. and Chen, Hao and Chang, Hao-Hsuan and Sehgal, Abhishek and Zhang, Jianzhong},
  title   = {Optimal preprocessing of {WiFi} {CSI} for sensing applications},
  journal = {IEEE Trans. Wireless Commun.},
  volume  = {23},
  number  = {9},
  pages   = {10820--10833},
  month   = sep,
  year    = {2024},
  doi     = {10.1109/TWC.2024.3376332}
}

@article{yang2023sensefi,
  author  = {Yang, Jianfei and Chen, Xinyan and Zou, Han and Lu, Chris Xiaoxuan and Wang, Dazhuo and Sun, Sumei and Xie, Lihua},
  title   = {{SenseFi}: A library and benchmark on deep-learning-empowered {WiFi} human sensing},
  journal = {Patterns},
  volume  = {4},
  number  = {3},
  pages   = {100703},
  year    = {2023},
  doi     = {10.1016/j.patter.2023.100703}
}

@article{zakeri2026freshness,
  title={Freshness-Aware Constrained Sensing-Aided Beam Prediction with Knowledge Distillation},
  author={Zakeri, Abolfazl and Nguyen, Nhan Thanh and Alkhateeb, Ahmed and Juntti, Markku},
  journal={arXiv preprint arXiv:2609.01225},
  year={Sep. 2026}
}

\end{document}